\documentclass[fleqn,usenatbib]{mnras}

\usepackage{newtxtext,newtxmath}

\usepackage[T1]{fontenc}

\DeclareRobustCommand{\VAN}[3]{#2}
\let\VANthebibliography\thebibliography
\def\thebibliography{\DeclareRobustCommand{\VAN}[3]{##3}\VANthebibliography}

\usepackage{graphicx}	% Including figure files
\usepackage{amsmath}	% Advanced maths commands
\usepackage[range-phrase=-, range-units=single, list-final-separator=\text{, and }, separate-uncertainty=true]{siunitx}

\title[Yaginuma, Taylor, \& Seligman]{Potential Thermal Profiles of The Third Interstellar Object 3I/ATLAS}

\author[Yaginuma, Taylor $\&$ Seligman]{
Atsuhiro Yaginuma,$^{1}$\thanks{E-mail: yaginuma@msu.edu}
Aster G. Taylor,$^{2}$\thanks{Fannie and John Hertz Foundation Fellow}
Darryl Z. Seligman$^{1}$
\\
$^{1}$Dept. of Physics and Astronomy, Michigan State University, East Lansing, MI 48824, USA\\
$^{2}$Dept. of Astronomy, University of Michigan, Ann Arbor, MI 48109, USA
}

\date{Accepted XXX. Received YYY; in original form ZZZ}

\pubyear{2025}

\begin{document}
\label{firstpage}
\pagerange{\pageref{firstpage}--\pageref{lastpage}}
\maketitle

\begin{abstract}

We investigate the thermal evolution of 3I/ATLAS, the third macroscopic interstellar object discovered on 2025 July 1. By comparing modeled thermal profiles with observations of volatile activity, it is possible to constrain bulk physical properties of a cometary nucleus. 3I/ATLAS is actively producing a variety of cometary volatiles. In this paper, we calculate one-dimensional thermal profiles of the third interstellar object 3I/ATLAS throughout its trajectory in an attempt to gain insight into its bulk properties based on measurements of its volatiles. Assuming a variety of typical comet and asteroid bulk geophysical properties such as heat capacities, densities, and conductivities, we calculate the radial thermal profile as a function of depth throughout the hyperbolic trajectory. The methods and code to generate the thermal profile are flexible for any hyperbolic or bound orbit. The thermal profiles are benchmarked to the nominal sublimation temperatures of H$_2$O, CO$_2$ and CO, but are still applicable to any volatile. Comparison between the modeled surface temperatures and the observed onset of H$_2$O activity near 3 au indicates that surface temperatures exceeding $\sim$150 K can only be achieved if the albedo is below 0.2. We therefore set the upper limit on the albedo of 3I/ATLAS to be 0.2.

\end{abstract}

\begin{keywords}
comets: general -- minor planets, asteroids: general -- radiation mechanisms: thermal
\end{keywords}

\section{Introduction}

The third macroscopic interstellar object 3I/ATLAS \citep{Denneau2025}, also known as C/2025N1 and originally called A11pl3Z,\footnote{\url{https://www.minorplanetcenter.net/mpec/K25/K25N12.html}} was discovered in the inner Solar System on 2025 July 1 by the Asteroid Terrestrial-impact Last Alert System (ATLAS; \citealt{Tonry2018a}). Astrometric fits to the ATLAS observations yield a perihelion distance $ q\sim 1.36$ au, orbital eccentricity of $e\sim 6.2$, inclination of $i\sim175^\circ$, and a hyperbolic velocity of $v_\infty\sim58$ km s$^{-1}$. The Galactic radiant and large excess speed of 3I/ATLAS imply age of 3--11 Gyr and an association with lower-metallicity stellar populations \citep{Taylor2025, Hopkins2025}.

Initial near-discovery and pre-discovery observations revealed that 3I/ATLAS was weakly active \citep{Jewitt2025a, Alarcon2025, Chandler2025,Frincke2025}. Its reflectance spectrum is reddened and largely featureless with an absolute magnitude of $H_V\sim12$ \citep{Seligman2025, Opitom2025}, and high resolution HST/WFC3 imaging constrains the nucleus radius to $\lesssim2.8$ km \citep{Jewitt2025b}. The visible and near-infrared colors match those of D-type asteroids and active Centaurs \citep{Belyakov2025, Kareta2025, Marcos2025, Tonry2025, Yang2025, Santana-Ros2025, Puzia2025, Beniyama2025}. Optical and near-infrared spectra show a broad \qty{2.0}{\micro\meter} absorption feature, consistent with $\sim\qty{30}{\percent}$ of the coma being composed of $\sim\qty{10}{\micro\meter}$ water-ice grains \citep{Yang2025}, and Neil Gehrels-Swift Observatory detections of OH confirm H$_2$O outgassing \citep{Xing2025}.

Time-series photometry yields a rotation period of 16.16$\pm$0.01 h and dust loss rates of 0.3--4.2 kg s$^{-1}$ \citep{Santana-Ros2025}. Shift-stacked images from NASA's Transiting Exoplanet Survey Satellite (TESS) \citep{Ricker2015} and pre-discovery observations from the Zwicky Transient Facility (ZTF) \citep{Bellm2019} point to low level activity at $\sim6.4$ au \citep{Feinstein2025, Martinez-Palomera2025, Ye2025, Farnham2025}. In addition, SOAR/Goodman \citep{Clemens2004} spectra exhibit a red continuum dominated by refractory organics with no CN, C$_2$, or CO$^{+}$ detected at 4.4 au \citep{Puzia2025}. 

The first two interstellar objects displayed sharply contrasting behavior. 1I/`Oumuamua showed no detectable dust activity in deep imaging observations \citep{Meech2017, Ye2017, Jewitt2017,Trilling2018}. Its trajectory deviated from purely gravitational motion and required a comet-like nongravitational acceleration \citep{Micheli2018} possibly explainable with outgassing \citep{Seligman2018, Seligman2020, Levine2021, Levine2022, Jackson2021, Bergner2023}. In contrast, 2I/Borisov behaved like a classical active comet, exhibiting a persistent dust coma and common gas species \citep{Jewitt2019, Fitzsimmons2019, Ye2019, McKay2020, Guzik2020, Hui2020, Kim2020, Cremonese2020, Yang2021}. However, the composition of 2I/Borisov was enriched in CO compared to H$_2$O \citep{Xing2020,Bodewits2020,Cordiner2020}. Both 1I/`Oumuamua and 2I/Borisov exhibited a lower excess velocity than that of 3I/ATLAS, implying younger kinematic ages, albeit with substantial uncertainties \citep{Mamajek2017, Gaidos2017, Feng2018, Fernandes2018, Hallatt2020, Hsieh2021, Taylor2025}. The diversity in activity, composition, and dynamics between the three interstellar objects motivates further investigation of these objects' physical properties.

Thermal modeling constrains an object's physical properties by connecting insolation, sublimation, and conduction to outgassing and other observable properties \citep{Deldo2015, Davidsson2014}. In practice, the absorbed energy is divided into thermal reradiation to space, latent heat consumed by sublimation, and conduction into the subsurface. Temperatures then evolve by diffusion, with the response set by the material’s capacity to store and conduct heat and further influenced by rotation, surface roughness, and the composition and distribution of volatile content such as H$_2$O, CO$_2$, and CO. \citep{Rozitis2012, Davidsson2014}. These properties control the diurnal temperature range and lag, the timing and location of sublimation, the ability of gas to loft dust, the thermal infrared output, and the magnitude and direction of any outgassing-driven acceleration.

For asteroids, thermal models can be fit to infrared data to obtain the effective diameter and geometric albedo of the object. When the spin and shape of the object are known, these fits will also calculate the object's thermal inertia and surface roughness, which together map to surface composition and heat storage \citep{Harris1998, Mueller1998, Rozitis2012, Deldo2015}. For comets, similar models couple surface temperatures to detected sublimation rates and enable inference of an object's active surface fraction \citep{Davidsson2014,Keller2015}.

In this paper, we apply a thermophysical framework to 3I/ATLAS to investigate how its surface and subsurface temperatures evolve along its hyperbolic trajectory. We explore a range of thermal properties and volatile compositions to identify the conditions under which sublimation of major species such as H$_2$O, CO$_2$, and CO occur. The resulting temperature profiles are compared with photometric and spectroscopic observations, including evidence of early activity and detections of outgassing. Finally, we discuss these results' implications for the surface and the post-perihelion evolution of 3I/ATLAS.

\begin{figure*}%[tbhp]
\centering
\includegraphics[width=1.\linewidth]{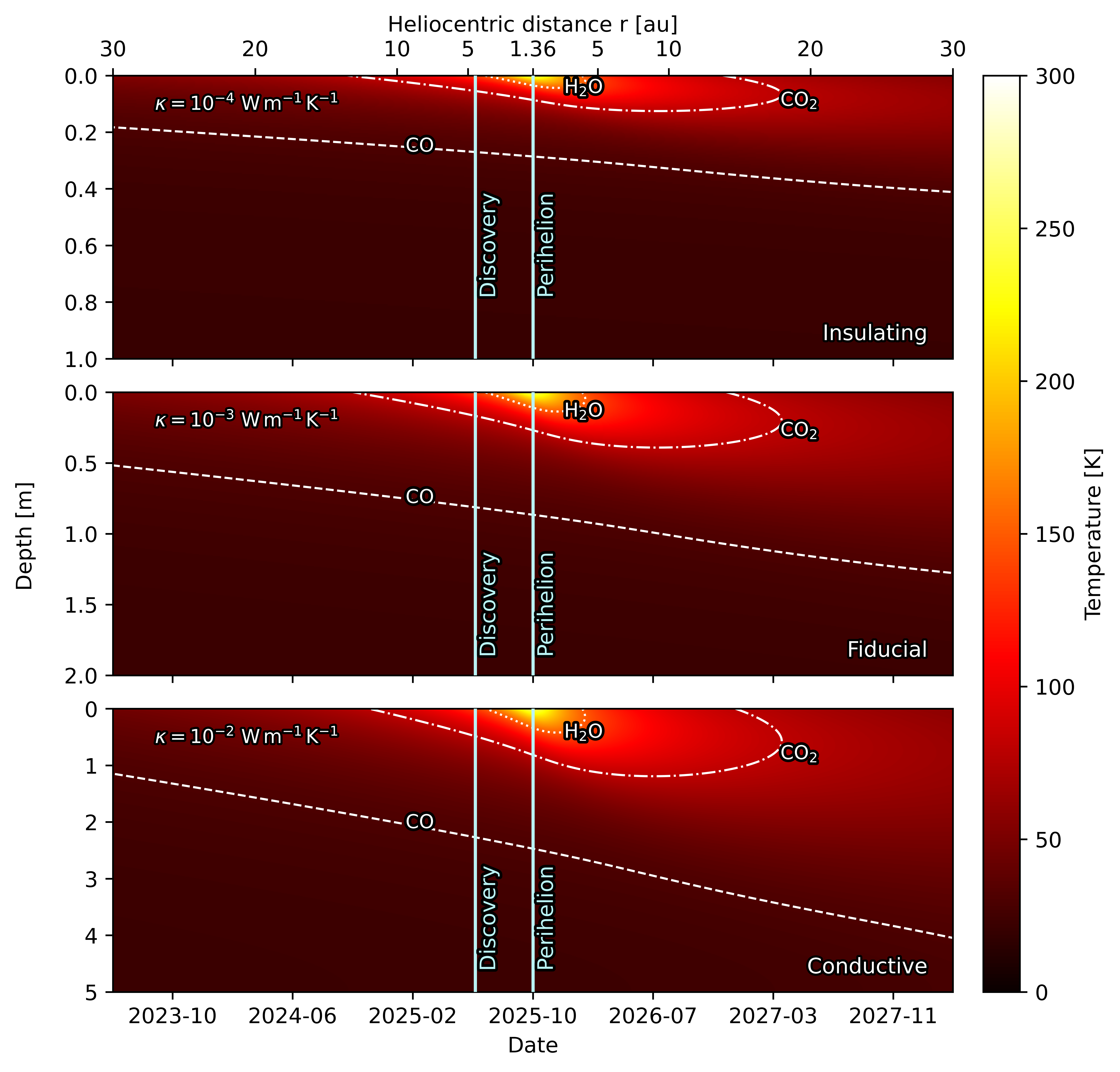}
\caption{
The temperature versus depth and time for $\kappa=\qtylist{e-4;e-3;e-2}{W\,m^{-1}\,K^{-1}}$ (top, middle, and bottom panels). The sublimation temperatures of CO (\qty{30}{K}, dash), CO$_2$ (\qty{80}{K}, dash-dot), and H$_2$O (\qty{150}{K}, dot) are shown as white contours. The depth of heat penetration and volatile activation strongly depends on the thermal conductivity $\kappa$. Note that the range of the $y$-axis is different for each panel.} 
\label{fig:thermal_H1}
\end{figure*}

\begin{figure}%[tbhp]
\centering
\includegraphics[width=1.\linewidth]{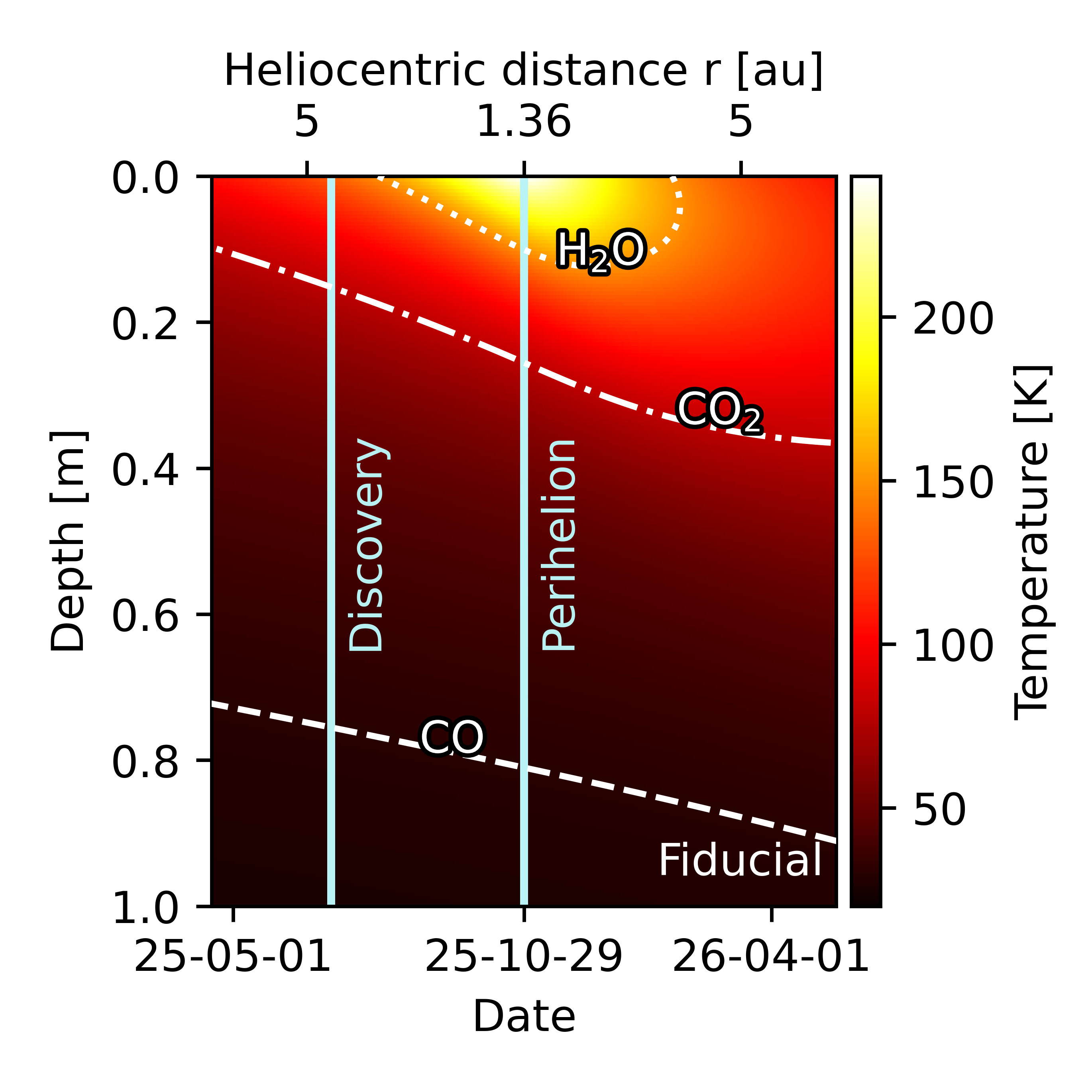}
\caption{Zoomed-in version of Figure \ref{fig:thermal_H1} closer to the discovery and perihelion dates. } 
\label{fig:thermal_zoomin}
\end{figure}

\begin{figure}%[tbhp]
\centering
\includegraphics[width=1.\linewidth]{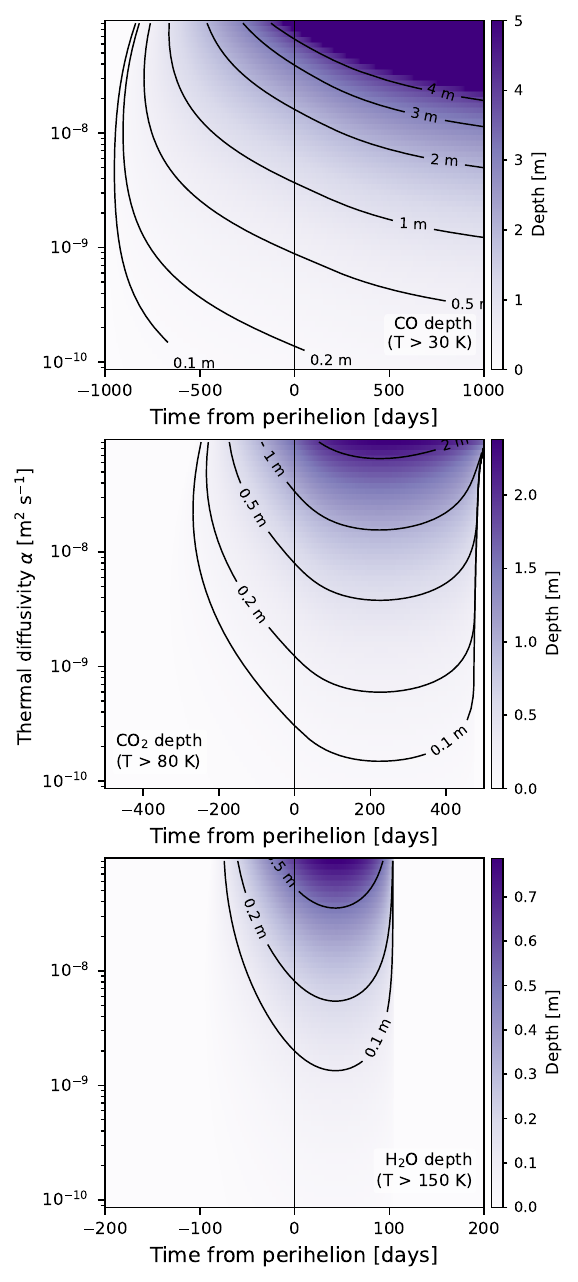}
\caption{
The maximum depth where a given species will sublimate as a function of time and the thermal diffusivity $\alpha$. The depth of volatile activity depends strongly on thermal diffusivity, with CO and CO$_2$ remaining active over a wide range of depths and times, while H$_2$O sublimation occurs only under warmer conditions near perihelion.
} 
\label{fig:thermal_diffusvity}
\end{figure}

\section{Methods} \label{sec:methods}

Our methodology follows the approach developed by \citet{Fitzsimmons2018} for the first interstellar object 1I/`Oumuamua.

We model heat transport in a one-dimensional Cartesian geometry, representing a vertical column normal to the surface. This approximation is valid when the modeled depth is much smaller than the object's radius. The computational domain extends to a maximum depth of \qty{10}{m}, beyond which the temperature gradient is assumed to be negligible. This depth is small compared to the $\mathcal{O}(\unit{km})$ expected size of 3I/ATLAS, so the Cartesian approximation holds. This slab model assumes a spherically symmetric nucleus and considers heat transfer by solid state conduction only. Because of this, our model does not include gas diffusion inside the nucleus or detailed interaction between the nucleus and the coma. More comprehensive thermophysical models show that gas diffusion and gas returning from the coma can change the surface energy balance and volatile outgassing, especially for material coming from deep layers (see, e.g., \citet{Prialnik2004, Davidsson2004} for a review). The temperature evolution is described by the heat diffusion equation
\begin{equation}
\rho C_p \frac{\partial T}{\partial t} = \frac{\partial}{\partial z} (\kappa \frac{\partial T}{\partial z}),
\end{equation}
where $T$ is the temperature at depth $z$ and time $t$. Here, $\kappa$ is the thermal conductivity, $\rho$ is the density, and $C_p$ is the heat capacity. Following \citet{Fitzsimmons2018}, we used comet-like constants $\rho=\qty{e3}{kg\, m^{-3}}$, and $C_p=\qty{550}{J\,kg^{-1}\,K^{-1}}$. We vary the thermal conductivity between $\kappa=\qtylist{e-4;e-3;e-2}{W\,m^{-1}\,K^{-1}}$, respectively corresponding to (i) insulating, (ii) fiducial, and (iii) conductive surface materials. These values correspond to thermal diffusivity of $\alpha  = \kappa/(\rho C_p) \simeq\qtylist{1.8e-10;1.8e-9;1.8e-8}{\meter^2/s}$.

The lower boundary condition is an insulating Dirichlet condition specified as $\partial T/\partial z=0$ at $z=z_{\rm max}=\qty{10}{m}$. At the surface ($z=0$), the absorbed solar flux balances the thermal reradiation and conductive heat flux into the subsurface so that
\begin{equation}
\kappa \frac{\partial T}{\partial z}\bigg|_{z=0}
= \epsilon \sigma T^4(0,t)
- \frac{f\,F_\odot(1-A)}{r^2(t)}.
\end{equation}
Here, $\epsilon$ is the bolometric emissivity, $\sigma$ is the Stefan-Boltzmann constant, $A$ is the Bond albedo, $F_{\odot}=\qty{1367}{W/m^2}$ is the solar constant at 1 au, and $f=1/{\pi}$ accounts for the rotationally averaged absorbed flux, which is appropriate for a rapidly rotating body with its spin axis roughly perpendicular to the direction of the Sun. The heliocentric distance $r(t)$ is computed from the object's orbital solution at each time step.

To solve these equations, the computational domain is divided into uniform layers of thickness $\Delta z=5$ mm and the temperature evolution is computed using an explicit finite-difference method, where spatial derivatives are evaluated with a central difference scheme. Since the radiation term depends on $T^4$, the Newton-Raphson method is used to iteratively solve the nonlinear surface boundary conditions at each time step. This ensures that the absorbed, emitted, and conducted fluxes are balanced to within numerical precision, conserving energy throughout the integration.

We initialized the simulation 6500 days before perihelion, when 3I/ATLAS was more than 100 au from the Sun. At such large distances, the solar flux is negligible, so the nucleus was assumed to be nearly isothermal. We set the initial temperature in all layers to the radiative equilibrium value,
\begin{equation}
T = \Bigl(\frac{fF_{\odot}(1-A)}{4\epsilon\sigma r^2}\Bigr)^{\frac{1}{4}},
\end{equation}
which is approximately 30 K at 100 au for $A=0.1$ and $\epsilon=0.95$. From this initial state, we integrated the thermal model forward in time over a total of 13000 days centered on the perihelion with timestep of 50 seconds. The temperature is recorded as a function of time and depth over this time frame.

Finally, we compared the calculated temperature with the approximate sublimation temperatures of the main volatiles (CO at \qty{30}{K}, CO$_2$ at \qty{80}{K}, and H$_2$O at \qty{150}{K}) to estimate the depth and timing of sublimation activity over the trajectory of 3I/ATLAS. Our thermal calculation does not include sublimation as an energy sink, which means the modeled surface temperatures are slightly higher than the real ones.

\section{Results}

\begin{figure}%[tbhp]
\centering

\includegraphics[width=1.\linewidth]{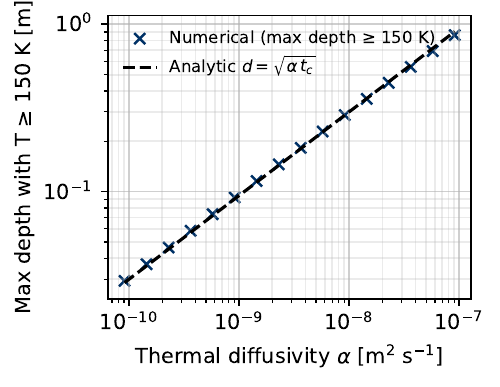}
\caption{The maximum depth where $T\geq\qty{150}{K}$ versus the thermal diffusivity. The numerical results (crosses) are shown versus an analytic scaling calculation (dashed line). These results indicate that the temperatures required for H$_2$O sublimation do not reach deeper than about 1 m for any thermal diffusivity. Subsurface water ice will be stable below this depth.} 
\label{fig:thermal_dif}
\end{figure}

\begin{figure}%[tbhp]
\centering
\includegraphics[width=1.\linewidth]{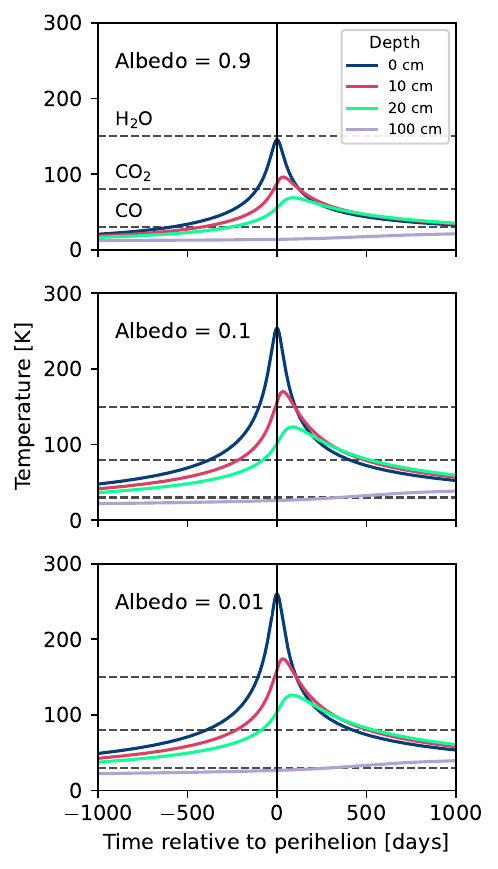}
\caption{
The temperature at series of depths versus time. The different panels show calculations with different albedoes. The surface reflectivity plays a key role in controlling thermal evolution, with high-albedo surfaces staying cooler and suppressing volatile release while low-albedo surfaces heat more efficiently, leading to deeper thermal penetration and stronger H$_2$O activity near perihelion. Here, we used comet-like constants and fixed thermal conductivity to $\kappa=\qty{e-3}{W\,m^{-1}\,K^{-1}}$.
} 
\label{fig:temp_depth_albedo}
\end{figure}

\begin{figure}%[tbhp]
\centering
\includegraphics[width=1.\linewidth]{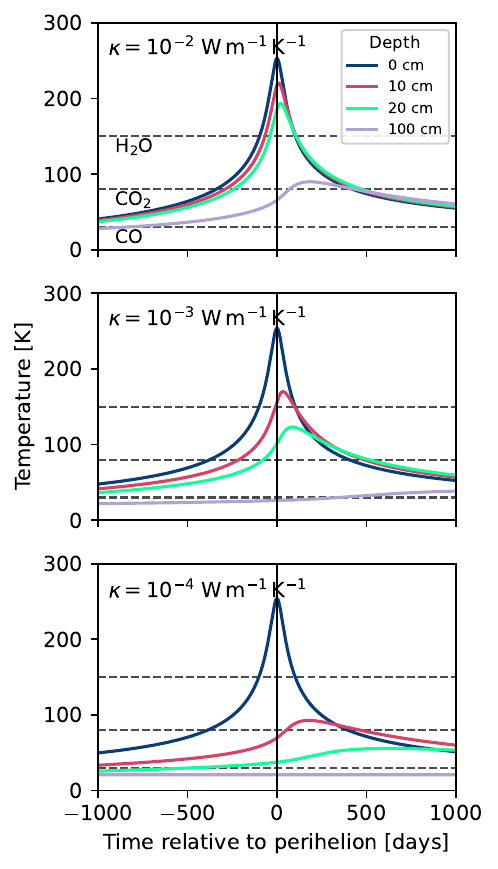}
\caption{
The temperature at series of depths versus time. The different panels show calculations with different thermal conductivities. Thermal conductivity governs how far heat can propagate below the surface, with larger $\kappa$ values enabling deeper sublimation fronts, while smaller $\kappa$ keeps heating confined to near-surface layers. Here, we used comet-like constants and fixed albedo to $A=0.1$.
} 
\label{fig:temp_depth_kappa}
\end{figure}

\subsection{Surface and Subsurface Temperature Evolution}

We first computed the surface and subsurface temperature evolution for the three cases described in Section \ref{sec:methods}, which span the range of expected surface thermophysical properties of comets and asteroids. Figure \ref{fig:thermal_H1} shows the modeled surface temperature of the body as a function of heliocentric distance and time for different thermal conductivity cases. The heat capacity and density are fixed at comet-like values, with an assumed albedo of $A=0.1$ and emissivity of $\epsilon=0.95$. The three contour lines show the nominal sublimation fronts of H$_2$O (150 K), CO$_2$ (80 K) and CO (30 K). These temperatures are determined from existing thermal evolution profiles \citep{Puzia2025, Levine2023, Fitzsimmons2018, Gasc2017}. Even under the most conductive conditions, H$_2$O sublimation front remains confined to within the top \qty{0.5}{m} and is active only for a short period near perihelion. In the fiducial case, H$_2$O sublimation occurs only within the upper \qty{0.2}{m} and for just a few weeks around closest approach to the Sun. In all cases, the onset of water sublimation occurs after the discovery.

In contrast, the sublimation temperatures of CO$_2$ and CO ices are reached at much larger heliocentric distances, so they would become thermally active earlier and over greater depth. The surface temperature reaches $\sim$80 K several hundred days before perihelion and reaches these temperatures at depths of \qty{1}{m} in the conductive case. On the other hand, surface CO (30 K) will sublimate throughout most of the orbit. 

Figure \ref{fig:thermal_zoomin} provides a zoomed view of the fiducial case around the perihelion. Under our applied conditions, H$_2$O begins to sublimate at a heliocentric distance of approximately 3.8 au. H$_2$O sublimation is limited to the upper tens of centimeters and lasts a few weeks past perihelion. CO$_2$ is sublimated at depths of 0.3--0.4 m and is activated earlier, while CO sublimation extends even deeper, remaining efficient down to roughly 1 m.

\subsection{Dependence on Thermal Diffusivity}

The dependence of volatile activity on thermal diffusivity is shown in Figure \ref{fig:thermal_diffusvity}. Each panel illustrates the depth at which the temperature exceeds the sublimation threshold for CO, CO$_2$, and H$_2$O as a function of time relative to perihelion and thermal diffusivity $\alpha$. The results reveal a clear hierarchy in the thermal response. With a sublimation temperature of about 30 K, CO can become active hundreds of days before perihelion even in highly insulating materials. Its sublimation front can reach meters of depth over a wide range of diffusivities and remains active long after perihelion. CO$_2$, which sublimates near 80 K, becomes active around 200 to 300 days before perihelion and stays active for a comparable time after. For conductive surfaces with high diffusivity, the sublimation front approaches \qty{1}{m}, while it remains shallower than 0.2 m for insulating surfaces. Finally, H$_2$O becomes active only within about 100 days of perihelion and only sublimates from the near-surface layers. Even in conductive materials, water sublimation typically occurs within about 0.5 m of the surface, suggesting that H$_2$O driven outgassing would be short lived and concentrated near perihelion.

To verify physical consistency, Figure \ref{fig:thermal_dif} shows a comparison between the numerically computed maximum depth where $T \ge 150$ K and an analytic scaling. The depth of activity can be analytically calculated as $d=\sqrt{\alpha t_c}$, where $t_c$ is the characteristic duration of the perihelion heating pulse \citep{Jewitt2017}. Using the timescale $t_c=9\times10^{6}$ seconds, which roughly corresponds to the onset of H$_2$O sublimation, the analytic curve closely matches the numerical results across the full range of $\alpha$ and demonstrates that the sublimation depth is well described by $\sqrt{\alpha}$. As the diffusivity increases, the depth of heating increases. For the fiducial case, the maximum depth heated above 180 K is about 0.2 m, and even for the most conductive case it remains below 0.7 m. This indicates that water ice sublimation would be limited to the uppermost layers and that water would not be significantly depleted at meter scale depths.

\subsection{Effects of Albedo and Thermal Conductivity}

Figures \ref{fig:temp_depth_albedo} and \ref{fig:temp_depth_kappa} explore the influence of surface reflectivity and thermal conductivity on the near-surface temperature evolution. Figure \ref{fig:temp_depth_albedo} shows the temperature versus depth for for a fixed conductivity $\kappa=10^{-3}$ W m$^{-1}$ K$^{-1}$ and variable albedo, while Figure \ref{fig:temp_depth_kappa} explores a fixed albedo $A=0.1$ and variable conductivity. Each panel shows the temperature at several depths as a function of time relative to perihelion, with horizontal dashed lines marking the sublimation temperatures of CO, CO$_2$, and H$_2$O. Surface reflectivity strongly influences the temperature distribution and the depth to which heat penetrates. High albedo surfaces reflect most incoming sunlight, keeping the surface relatively cool and limiting heating of the subsurface. Under these conditions, water ice remains stable and only CO and CO$_2$ sublimate at all, and then only near the surface and close to perihelion. Moderate albedo surfaces absorb more sunlight, leading to higher surface temperatures and deeper heat penetration. CO$_2$ can sublimate deeper below the surface, and water ice can begin to sublimate near the surface. Low albedo surfaces absorb almost all the incident radiation, producing high surface temperatures and allowing the thermal wave to penetrate deeply. Even a meter layer warms significantly, and CO and CO$_2$ sublimation occurs over larger volumes and for longer durations. In all cases, deeper layers heat more slowly, reach lower peak temperatures, and exhibit a delayed response relative to the surface due to thermal inertia.

\begin{table*}
\centering
\caption{Summary of volatile detections from 3I/ATLAS observations. Listed production rates are based on published measurements or reported upper limits. Columns correspond to telescope, heliocentric distance measured in au, species, production rate in 10$^{26}$ molecules s$^{-1}$, and the relevant citation.}
\begin{tabular}{@{}llllll@{}}
\hline
Telescope / Instrument &  Distance  & Species  & Production Rate & Reference \\
 &   [AU] &   & [10$^{26}$ molecules s$^{-1}$] &  \\
\hline
JCMT & 2.13 & HCN & 0.4 $\pm$ 0.17 &\citet{Coulson2025} \\
Keck-II / KCWI & 2.75 & CN & 0.00091 & \citet{Hoogendam2025} \\
VLT / UVES + X-shooter & 2.85 & CN & 0.00407 & \citet{Rahatgaonkar2025} \\
VLT / UVES + X-shooter & 2.85 & Ni & 0.000468 [atom s$^{-1}$] & \citet{Rahatgaonkar2025} \\
MDM Observatory  & 3.2–2.9 & CN & 0.06 & \citet{Salazar_Manzano2025} \\
SPHEREx + NASA IRTF/SpeX & 3.3-3.1 & CO$_2$ & 9.4 & \citet{Lisse2025} \\
SPHEREx + NASA IRTF/SpeX & 3.3-3.1 & H$_2$O & 1.5 & \citet{Lisse2025} \\
SPHEREx + NASA IRTF/SpeX & 3.3-3.1 & CO &  2.8 & \citet{Lisse2025} \\
JWST NIRSpec & 3.32 & CO$_2$ & 17 $\pm$ 0.1 & \citet{Cordiber2025} \\
JWST NIRSpec & 3.32 & H$_2$O & 2.23 $\pm$ 0.08 & \citet{Cordiber2025} \\
JWST NIRSpec & 3.32 & CO & 3.7 $\pm$ 0.2 & \citet{Cordiber2025} \\
JWST NIRSpec & 3.32 & OSC & 0.017 & \citet{Cordiber2025} \\
Swift UVOT & 3.51 & H$_2$O (via OH) & 13.5 & \citet{Xing2025} \\
\hline
\end{tabular}
\label{table:observations}
\end{table*}

Thermal conductivity also plays an important role in shaping the temperature structure. High-conductivity materials transport heat efficiently into the subsurface, resulting in deeper warming and longer thermal retention after perihelion. Under these conditions, sublimation thresholds for CO$_2$ and even H$_2$O can be reached at greater depths. In the intermediate case, the behavior is balanced, with significant subsurface warming and a noticeable delay in peak temperature with depth. In the low conductivity case, heat does not travel far, and the surface experiences a sharp temperature spike near perihelion while deeper layers remain cold and stable. CO sublimation depth reaches beyond 1 meter for largest $\kappa$ while CO$_2$ and H$_2$O under 1 meter from the surface.

\subsection{Other Models and Thermal Behavior}

While our figures show the temperature evolution for a range of representative albedo and thermal conductivity values, Figure 11 in \citet{Chandler2025} present a thermophysical model with Bond albedo of $A=0.05$ and emissivity of $\epsilon=0.9$. For comparable parameters, our temperature profiles exhibit similar behavior, indicating good consistency between our thermal evolution results and those of their independent model. \citet{Puzia2025} also present a complementary approach in their Figure 2. Their model assumes a fixed Bond albedo of $A=0.04$, a bolometric emissivity of $\epsilon=0.95$, and a heliocentric distance–dependent solar flux. It also starts from an initial uniform internal temperature representative of the interstellar medium, about 10 K, at \qty{e5}{d} before perihelion. Despite using different assumptions, their results show a similar thermal behavior. The deeper layers heat up much more slowly than the surface, the amount of heat that reaches below depends strongly on the thermal properties, and the sublimation fronts change in a similar way. This agreement supports the reliability of our results and shows that both models capture the main physics of how heat moves and how volatiles become active in 3I/ATLAS.

Overall, the results show that the thermal behavior and volatile activity of 3I/ATLAS depend strongly on its physical properties. CO is likely to drive activity during most of the orbit, while CO$_2$ contributes both before and after perihelion. H$_2$O activity occurs only near perihelion and is limited to shallow layers, with its extent controlled by albedo and thermal conductivity.

\section{Discussion}

Understanding the physical nature of rare objects like 3I/ATLAS requires connecting observational evidence with thermal evolution. Measurements of gas production, spectral features, and dust behavior provide direct insights into the composition and current activity of the nucleus. At the same time, the thermal profile predicts when and where volatile species become active, how heat is transported into the subsurface, and the object's physical properties. Combining these models with observations provide greater understanding of the composition and evolution of 3I/ATLAS and enable comparison to both Solar System and other interstellar objects.

Recent detections of volatile species provide valuable constraints on the thermophysical state of 3I/ATLAS, as summarized in Table \ref{table:observations}. Ultraviolet observations with Swift revealed OH emission corresponding to a water production rate of $Q(\text{H}_2\text{O})\simeq\qty{1.36\pm0.35e27}{molecule/s}$ at the heliocentric distance of 2.9 au \citep{Xing2025}. SPHEREx and NASA IRTF/SpeX observations further detected strong CO$_2$ emission ($Q(\text{CO}_2)\simeq\qty{9.4e26}{molecule/s}$) and evidence of extended water ice absorption features \citep{Lisse2025}. These detections show that both CO and CO$_2$ became active well before 3I/ATLAS reaches perihelion and are likely the main drivers of activity during its inbound phase. This early CO$_2$-driven activity and steady brightening agree with the pre-perihelion observations by \citet{Jewitt2025c} which found that 3I/ATLAS brightened roughly as $r_H ^{-3.8}$, consistent with CO$_2$ sublimation dominating between 4 and 2 au. \citet{Keto2025} also showed that the appearance and fading of the sunward anti-tail can be explained by CO$_2$ outgassing that lifts short-lived H$_2$O ice grains, which sublimate near 3–4 au.

The most distant water detection, inferred from OH emission, was made by Swift, which measured $Q(\text{H}_2\text{O})\simeq\qty{1.35\pm0.27e27}{molecule/s}$ at an inbound heliocentric distance of 3.51 au \citep{Xing2025}. Assuming that this detection marks the onset of water sublimation, our thermal evolution profile constrains the surface to reach temperatures of about 150 K at 3.51 au, requiring a Bond albedo of $A\leq0.2$. If water emission were detected around 3.8 au, it would imply an albedo lower than 0.1. However, such a scenario is unlikely, since at that distance water-ice sublimation is expected to be weak and most of the observed activity is instead driven by CO$_2$. This provides a quantitative upper limit on the reflectivity of the surface and directly connects the observed activity to the modeled thermal environment.

Additional detections of other species were also reported. Observations with the MDM Observatory detected CN gas with a production rate of $Q(\text{CN}) \simeq \qty{6e24}{molecule/s}$ \citep{Salazar_Manzano2025}, while observations with JCMT measured HCN at \qty{4.0e25}{molecule/s} \citep{Coulson2025}. VLT/UVES and X-shooter revealed faint CN (\qty{4.07e23}{molecule/s}) and Ni (\qty{4.68e22}{atom/s}) emission between 4.4 and 2.85 au \citep{Rahatgaonkar2025}, while Keck-II/KCWI observations measured CN at \qty{9.1e22}{molecule/s} at 2.75 au \citep{Hoogendam2025}. These detections show that significant amounts of gas are released between 2.1 and 3.2 au, suggesting that volatile materials can sublimate from relatively shallow layers without the need for deep heating.

The gas production rate also provide constraints on nucleus size and activity level. Assuming an infrared emissivity of 0.95, albedo of 0.04, and a nucleus radius of 2.8 km \citep{Jewitt2025b} and using the model of \citet{Cowan1979}, the minimum active fractional areas are $> \qty{2.6}{\percent}$ for CO$_2$, $> \qty{0.14}{\percent}$, and $>\qty{0.62}{\percent}$ for H$_2$O \citep{Cordiber2025}. If the observed water outgassing near 3 au is driven by equilibrium sublimation, the minimum required active area is 7.8 km$^{2}$ \citep{Xing2025}. Then combined with the Hubble Space Telescope upper limit on the nucleus size ($R<\qty{2.8}{km}$; \citealt{Jewitt2025b}; 3.8 au before perihelion) this implies that more than \qty{8}{\percent} of the surface must be actively sublimating. 
% \textbf{The Hubble Space Telescope observations of \citet{Jewitt2025b} were made at a heliocentric distance of 3.8 au before perihelion.} 
This is greater than the \qty{3}{\percent} active surface fraction of 1I/`Oumuamua \citep{Hui2019}. This large active fraction suggests that the surface is very porous and contains fine-grained ice, making it easy for sunlight to reach and sublimate the ice.

All of the evidence together suggests that 3I/ATLAS is a volatile-rich interstellar object with a low to moderate reflectivity, albedo $A<0.2$. While our model reproduces the observed onset of volatile activity under these conditions, it does not explicitly include sublimation as an energy sink, which may modify these results.
These characteristics nevertheless suggest that 3I/ATLAS has preserved much of its original volatile material since it formed, offering a rare opportunity to study the composition and thermal evolution of material that originated in another planetary system.

\section*{Acknowledgments}

We thank the anonymous reviewer for insightful comments and constructive suggestions that strengthened the scientific content of this manuscript.

We thank Tessa Frincke and Luis Salazar Manzano for helpful conversations.

A.Y. acknowledges support from the Michigan State University Honors College Gillette Research Scholar Fund.
A.G.T. acknowledges support from the Fannie and John Hertz Foundation and the University of Michigan's Rackham Merit Fellowship Program. 

\section{Data and Software Availability}

The Python script used to generate the thermal evolution heatmap is available at https://github.com/yatsu77/3I-ATLAS-thermal-profile.git.

%%%%%%%%%%%%%%%%%%%%%%%%%%%%%%%%%%%%%%%%%%%%%%%%%%

\bibliographystyle{mnras}
\bibliography{example} % if your bibtex file is called example.bib

% Don't change these lines
\bsp	% typesetting comment
\label{lastpage}
\end{document}